\documentclass[prl,twocolumn,showpacs,preprintnumbers,amsmath,amssymb,superscriptaddress]{revtex4}

\usepackage{graphicx}
\usepackage[ansinew]{inputenc}
\usepackage{array}
\usepackage{color}
\usepackage{amsmath}
\usepackage{amsxtra}
\usepackage{amstext}
\usepackage{amssymb}
\usepackage{latexsym}

\usepackage{sidecap}
\usepackage{floatflt}

\begin{document}

\title{Optically controlled orbital angular momentum generation in a polaritonic quantum fluid}

\author{S.M.H. Luk}
\affiliation{Department of Physics, University of Arizona, Tucson, AZ 85721, USA}

\author{N.H. Kwong}
\affiliation{College of Optical Sciences, University of Arizona, Tucson, AZ 85721, USA}

\author{P. Lewandowski}
\affiliation{Physics Department and Center for Optoelectronics and Photonics Paderborn (CeOPP), Universit\"at Paderborn, Warburger Strasse 100, 33098 Paderborn, Germany}

\author{S. Schumacher}
\affiliation{Physics Department and Center for Optoelectronics and Photonics Paderborn (CeOPP), Universit\"at Paderborn, Warburger Strasse 100, 33098 Paderborn, Germany}
\affiliation{College of Optical Sciences, University of Arizona, Tucson, AZ 85721, USA}

\author{R. Binder}
\affiliation{College of Optical Sciences, University of Arizona, Tucson, AZ 85721, USA}
\affiliation{Department of Physics, University of Arizona, Tucson, AZ 85721, USA}

\date{\today}

\begin{abstract}
Applications of the orbital angular momentum (OAM) of light range from the next generation of optical communication systems to optical imaging
and optical manipulation of particles. Here we propose a micron-sized semiconductor source which emits light with pre-defined OAM components.
 This source is based on a polaritonic quantum fluid. We show how in this system modulational instabilities can be controlled and
 harnessed for the
 spontaneous formation of OAM components not present in the pump laser source.
 Once created, the OAM states exhibit exotic flow patterns in the quantum fluid, characterized by  generation-annihilation pairs.
 These can only occur in open systems, not in equilibrium condensates,
  in
contrast to well-established vortex-antivortex pairs.
\end{abstract}



\maketitle



The physics of orbital angular momentum of light (OAM) has attracted considerable attention (\cite{allen-etal.03,torres-torner.11,mansuripur.17} and references therein). The interest in the physics of OAM extends beyond the characterization and preparation of light beams with non-zero OAM, and includes such topics as rotational frequency shifts \cite{bialynicki-birula-etal.97,courtial-etal.98}, detailed analysis of the vortex physics \cite{swartzlander.07}, the physics of OAM in second-harmonic generation \cite{courtial-etal.97}, optical solitons with non-zero OAM \cite{firth-skryabin.97}, transfer of OAM from pump to down-converted beam \cite{martinelli-etal.04}, data transmission using OAM multiplexing \cite{wang-etal.12},
and quantum optical aspects such as entanglement \cite{mair-etal.01}. In addition, manipulation of
the circular polarization and
OAM states in liquid crystals designed to provide effective spin-orbit interaction has recently been reported \cite{loussert-etal.13}.

Unrelated to the manipulation of OAM is the vast body of research on exciton polaritons in semiconductor microcavities. Here, being part of the polaritons, otherwise non-interacting photons experience an effective interactions through the polaritons' excitonic component. This combines the advantages of nonlinear systems with the ease of measuring optical beams. Polaritons form a quantum fluid, and prominent example of observed phenomena include parametric amplification \cite{Savvidis2000} and Bose condensation (\cite{deng-etal.10,snoke-littlewood.10} and references therein).

\begin{figure}
\centerline{\includegraphics[scale=0.45,angle=-00,trim=00 150 70 120,clip=true]{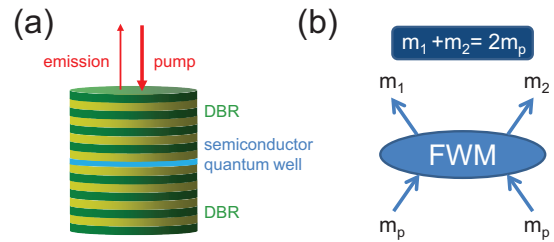}}
\caption{
 (a) Sketch of cylindrical semiconductor microcavity with distributed Bragg reflector (DBR) mirrors and one (or several) semiconductor quantum wells. (b) Schematic of a four-wave mixing (FWM) scattering process involving two incoming polaritons with orbital angular momentum  $m_p$ and two outgoing polaritons (spontaneously created if FWM exhibits instability) with $m_1$ and $m_2$.}
\label{fig:experimental-setup}
\end{figure}

The question then arises whether the relatively strong interaction between polaritons can be used to manipulate, in a well-controlled fashion, the orbital and/or spin angular momentum of polaritons (and thus the light field emitted from the cavity). For example, is it possible to use a beam with OAM of $m_p$  and create additional OAM contributions, say two components of OAM  $m_1$ and $m_2$, and to use the light beam characteristics of frequency and intensity to control $m_1$ and $m_2$?
The fact that rotationally symmetric states can be unstable under sufficiently large interactions is well known, and examples include spatial pattern formation
in chemical reactions and  biological process of morphogenesis \cite{turing52,ball99},
patterns in nonlinear optical solid-state and gaseous systems \cite{staliunas98,Kheradmand2008,Dawes2005}, and pattern \cite{ardizzone-etal.13} and vortex formation in atomic and polaritonic quantum fluids \cite{weiler-etal.08,keeling-berloff.08,dominici-etal.15}. However, the mere breaking of rotational symmetry does not answer the question how one could design a nonlinear system that would perform similarly to the linear system in \cite{loussert-etal.13}, but with the benefit of all-optical control of its output. In the following, we show that polaritonic interactions can be harnessed to create angular momentum states. The underlying physics is that of four-wave mixing instabilities, which is also used to create the above-mentioned optical patterns. However, in conventional pattern formation {\em linear momentum} (or wave vector) states become unstable and a state with wave vector $  \textbf{k}_p $ can generate waves with $  \textbf{k}_1 $ and $  \textbf{k}_2 $, and translational invariance dictates
 $ 2  \textbf{k}_p = \textbf{k}_1 + \textbf{k}_2   $.
 Many nonlinear systems allow only for a limited number of patterns, for example, in the case of waves vector instabilities, stripes (also called rolls) and hexagonal patterns \cite{newell-etal.93}.

In the present case, we consider a system that is not translationally invariant.
We show that in a cylindrical cavity with finite radius pumped with polaritons of OAM $m_p$, four-wave mixing instabilities lead to the spontaneous pairwise formation of OAM states  with $m_1$ and $m_2$, subject to the condition of angular momentum conservation $2 m_p = m_1 + m_2$,
Fig.\ \ref{fig:experimental-setup}. In analogy to the conventional patterns, we are interested in creating only a few stationary (at least on the nanosecond time scale considered here) OAM states.
In contrast to \cite{krizhanovskii-etal.10,tosi-etal.11}
we are not creating vortex-antivortex pairs, i.e.\ regions of oppositely rotating flow patterns that correspond to opposite
signs of  $ \vec{\nabla} \times \textbf{j}$ (where $ \textbf{j}$ is the particle current).
 Instead, we are creating generation-annihilation
  pairs, i.e.\ regions that
 generate and destroy polaritons corresponding to opposite signs of $ \vec{\nabla} \cdot \textbf{j}(\textbf{r})$.
 In contrast to vortex-antivortex pairs, which can exist in closed systems and condensates in thermal equilibrium,
 generation-annihilation pairs can only exist in open systems.
  Moreover, we are not using seed beams (called imprint beams in \cite{krizhanovskii-etal.10}) to trigger the instability; our instability
 is triggered by fluctuations and self-sustained, requiring only the pump to be present.

In semiconductor microcavities, there is also a spin-orbit interaction, giving rise, for example, to a polaritonic Hall effect, called the optical spin Hall effect \cite{leyder-etal.07}. A conceptually simple and robust all-optical control scheme of this effect has been demonstrated in \cite{lafont-etal.17}. The question then arises whether the spin-orbit interaction,
caused by the splitting of transverse-electric (TE) and transverse-magnetic (TM) modes,
 renders the pure orbital angular momentum control impossible, as orbital and spin angular momentum are not independently conserved. For realistic values of the TE-TM splitting,
 we find the effect of the spin-orbit interaction (SOI) to be unimportant and, for clarity, omit it in the following discussion (results including SOI are in the
 Supplemental Material).

We treat the polariton system as a two-dimensional coherent field moving in the cavity plane (Fig. \ref{fig:experimental-setup}a). The field's time evolution is governed in our model by a spinor Gross-Pitaevskii equation:
\begin{align}
i \hbar \frac {\partial \Psi^{\pm}} {\partial t} &= \left[ \hbar \omega_0 - \frac {\hbar^2} {2 M} \nabla^2 + V - i \gamma \right] \Psi^{\pm} + T^{++} | \Psi^{\pm} |^2 \Psi^{\pm} \notag \\
&\quad + T^{+-} | \Psi^{\mp} |^2 \Psi^{\pm} + S^{\pm}\label{GP.equ}
\end{align}
Here $\Psi^{\sigma} ( \textbf{r} , t )$ is the polariton field, with $\textbf{r} = ( x , y )$ and $\sigma = + , -$ denoting the spin (polarization) state relative to the $z$ axis (normal to the cavity's plane). The source term $S^{\pm} (  \textbf{r} , t )$ plays the role of an external pump, and $\gamma / \hbar$ is the combined rate of dissipative and radiative losses. $\hbar \omega_0$ is the minimum  polariton energy, $M$  the polariton mass,  and $T^{++}$ and $T^{+-}$ are parameters representing the scattering amplitudes between two polaritons with parallel and anti-parallel spins respectively.
For analytical advantages, we formulate Eq.\ (\ref{GP.equ}) in the angular momentum basis. The fields are expanded as
$\Psi^{\pm} (  \textbf{r} , t ) = \sum\limits_{m \in \mathbb{Z}} \psi^{\pm}_{m} ( r , t ) e^{i m \phi - i \omega_p t}$  (similar for $S^{\pm} (  \textbf{r} , t )$)
 %
 %
 where $( r , \phi )$ are spatial polar coordinates, and $\omega_p$ is the pump's center frequency.
 Eq.\ (\ref{GP.equ}) yields the equations of motion of the radial components as
\begin{align}
i \hbar \frac {\partial} {\partial t} \psi^{\pm}_{m} (r, t) &= \left[ - \mathcal{L}_{m} + V(r) \right] \psi^{\pm}_{m} (r , t) + s^{\pm}_{m} (r , t) \label{radial-GP.equ} \\
&\quad + \sum\limits_{m' m'' m'''} \delta_{m + m''' , m' + m''} \notag \\
&\quad \quad \times \left[ T^{++} \psi^{\pm \ast}_{m'''} (r , t) \psi^{\pm}_{m'} (r , t) \psi^{\pm}_{m''} (r , t) \right. \notag \\
&\quad \quad + \left. T^{+-} \psi^{\mp \ast}_{m'''} (r , t) \psi^{\mp}_{m'} (r , t) \psi^{\pm}_{m''} (r , t) \right] \notag
\end{align}
where
$\mathcal{L}_{m} = \frac {\hbar^2} {2 M} \left( \frac {\partial^2} {\partial r^2} + \frac {1} {r} \frac {\partial} {\partial r} + k^2_p - \frac {m^2} {r^2} \right) + i \gamma$ and $\frac {\hbar^2 k^2_p} {2 M} = \Delta_p = \hbar ( \omega_p - \omega_0 )$.
$V ( r )$ is a circularly symmetric potential confining the polaritons to a region $r \leq R$, requiring $\psi^{\pm}_{m} (R) = 0$.
 The cubic nonlinear terms represent two-polariton scattering between angular momentum states.

Eq.\ (\ref{radial-GP.equ}) is solved explicitly in time domain simulations.
The pump is taken to be $(+)$ polarized with an OAM equal to $m_p$: $s^{\sigma}_{m} = \delta_{m m_p} \delta_{\sigma +} S(r)$.
Since the system's setup is circularly symmetric, in the absence of instabilities, the excited coherent polariton keeps the pump's OAM $m_p$. (Incoherent polaritons carrying other $m$'s are produced by scattering out of $\Psi^{\pm}$, which is considered as part of the loss $\gamma$ in Eq. (\ref{radial-GP.equ})). The nonlinear terms, however, include four-wave mixing processes that may drive rotational modulational instabilities of the pumped polariton field, creating new angular momentum components. One such process, shown in Fig. \ref{fig:experimental-setup}b, is a scattering of two polaritons in the pump mode into two modes with OAM $m_1$, $m_2$, the values of which are restricted by OAM conservation $m_1 + m_2 = 2 m_p$. Under favorable conditions, this scattering triggers the instability, which can result in optical parametric oscillation (OPO), by enabling mutually reinforcing growth in the polariton components in modes $m_1$ and $m_2$. This is the rotational analog of the familiar translational FWM instability where two plane waves of equal and opposite linear momenta arise spontaneously out of a uniform field.

A challenge common to many nonlinear systems is to find stationary pattern-like solutions to the corresponding nonlinear equation. Performing large-scale numerical solutions of the nonlinear equations and scanning all control parameters is often prohibitively numerically expensive. Instead, one often performs a linear stability analysis (LSA). This helps identifying modes that are linearly unstable, but it does not guarantee that those modes uniquely identify the  emerging stationary patterns \cite{newell-etal.93}. In the following, we therefore combine a simplified LSA, which allows us to obtain qualitative insight into possible instability scenarios, and then use numerical solutions of the full nonlinear equation to study the system for parameter values close to those suggested by LSA results. In our LSA,
we assume the polariton component in the pump mode
to be a constant, independent of $r$, which we  denote by  $| {\bar \psi}^{+}_{p} |^2$
and the pump is monochromatic.
Below, in the full numerical solution, a source $S(r,t)$  excites a $ \psi^{+}_{m_p}(r)$ that is almost r-independent but drops to zero close to $R$. There, the LSA results are a
useful starting point for a numerical search for the desired instabilities.
Within the LSA, $| {\bar \psi}^{+}_{p} |^2$
is determined by $S$ (now a number, not a function $S(r,t)$) through Eq.\ (\ref{radial-GP.equ}). The rotational stability of this steady state is examined by linearizing Eq.\ (\ref{radial-GP.equ}) in fluctuations in mode pairs $( m_1 , m_2 )$ satisfying $m_1 + m_2 = 2 m_p$ and solving the attendant eigenvalue problem. Details of the LSA can be found in the Supplementary Material. The stability eigenvalue $\lambda$ (the linearized fluctuations are proportional to $e^{- i \lambda t}$ so that ${\rm Im} \lambda > 0$ implies instability), for $m_p = 0$, is given by
\begin{equation}\label{eigenvalues-finite-r.equ}
\hbar \lambda = - i \gamma \pm \sqrt{  \left(  \varepsilon_{m n} - 2 T^{++} | {\bar \psi}^{+}_{p} |^2 \right)^2 - \left( T^{++} | {\bar \psi}^{+}_{p} |^2 \right)^2}
\end{equation}
where
$\varepsilon_{m n} = \frac{ \hbar^2 } {2M} \left[ k^2_p -  (\alpha_{m n} / R )^2  \right]$ with $\alpha_{m n}$  the $n$-th zero, not counting the origin, of the Bessel function $J_m$.

\begin{figure}
\centerline{\includegraphics[scale=0.65,angle=-00, trim=0 0 0 0,clip=true]{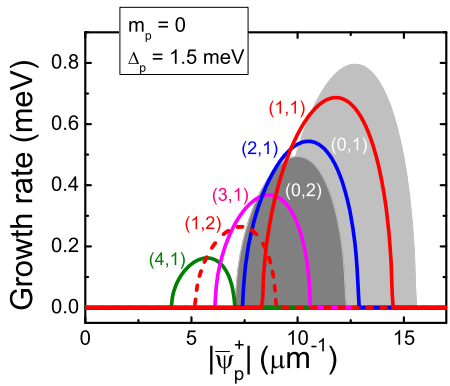}}
\caption{
The maximum growth
 rate $\Lambda$
vs.\  $|\bar{\psi}^{+}_{p}|$.
The brackets $(|m|,n)$ indicate the OAM and n-th zero, respectively.
Here, $m = \pm 1, \pm 2, \pm 3, \pm4$ OAM modes have a positive growth rate.
The growth rate of $\psi^{+}_{m_p}$ ($m_p = 0$) with $n=1$ and $n=2$ are shown as grey areas ($n > 2$ are stable).
We see a range of $|\bar{\psi}^{+}_{p}|$ values (just below 5 $\mu$m$^{-1}$) where (4,1) exhibits instability while the pump is stable
(cf.\ Fig. \protect\ref{fig:bargraphintensity}b)
}
\label{fig:LSA15}
\end{figure}

Denoting the imaginary part of the square root in Eq.\ (\ref{eigenvalues-finite-r.equ}) by $\Lambda$, we show in Fig.~\ref{fig:LSA15}
the maximum growth rate
 $\Lambda = {\rm Im}(\hbar \lambda) + \gamma$
 plotted against the pump-mode polariton component $|{\bar \psi}^{+}_{p} |$  for pump OAM $m_p = 0$ and detuning $\Delta_p = 1.5~\text{meV}$. The physical parameters are
 $M = 7.2\times 10^{-5} m_0$ ($m_0$=free electron mass),
  $T^{++} = 5.69\times10^{-3} \text{meV}\mu\text{m}^2$, and $R = 5 \, \mu\text{m}$.
With these parameters, the OAM modes with $| m | = 1 , \cdots , 4$ have positive growth rate over various ranges of $|{\bar \psi}^{+}_{p} |$ values. However fluctuations in the pump OAM channel may also destabilize the pumped steady state. This instability is marked for two $m = 0$ modes by the grey areas in Fig.~\ref{fig:LSA15}. The overlap of the instability ranges of the modes with $m = 0$ and those with $| m | = 1,\cdots, 3$ implies that the latter are not effective instability channels. Hence, only the $| m | = 4$ modes are expected to carry instability growth (at least initially) in this example. This kind of stability information is useful for determining parameters for simulations and experiments of OAM mode creations.

\begin{figure}
\centerline{\includegraphics[scale=1.1,angle=-00,trim=0 0 20 0,clip=true]{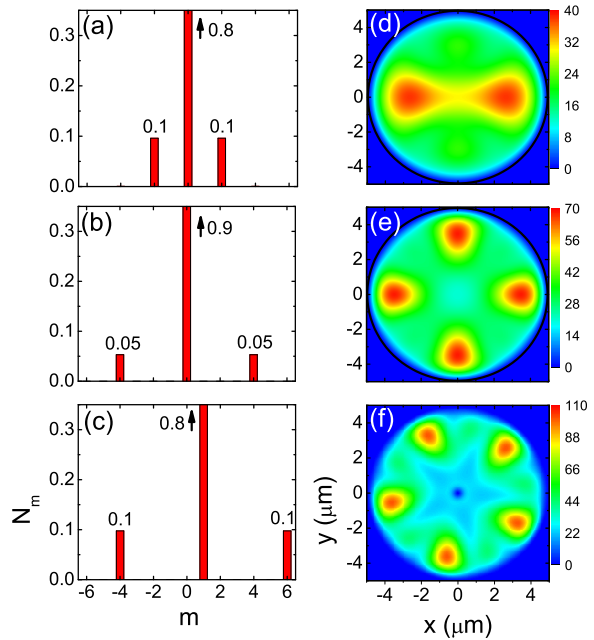}}
\caption{Left:
 The polariton density OAM fractions $N_m$
 under self-sustained optical parametric oscillation conditions (rounded values indicated in figure).
 Spontaneously created modes have OAM (a) $m_1=+2$ and $m_2=-2$, (b) $m_1=+4$ and $m_2=-4$, and (c) $m_1 = 6$ and $m_2 = -4$.
 Parameters:
 (a) $m_p = 0$, $\Delta_p = 0.7$meV, $\gamma = 0.05$meV,  $| \bar{ \psi}^{+}_{p} | = 4.0~\mu\text{m}^{-1}$;
 (b) $m_p = 0$,  $\Delta_p = 1.5$meV, $\gamma = 0.15$meV,  $| \bar{ \psi}^{+}_{p} | = 5.5~\mu\text{m}^{-1}$;
 (c) $m_p = 1$, $\Delta_p = 2.0$meV $\gamma = 0.15$meV,  $| \bar{ \psi}^{+}_{p} | = 5.7~\mu\text{m}^{-1}$.
Right:
(d)-(f) Corresponding real-space distribution  of polariton density $| \Psi^{+}(\textbf{r}) |^2 $ (units of $\mu$m$^{-2}$).
 }
\label{fig:bargraphintensity}
\end{figure}

The time evolution of the polariton field is followed beyond the linear instability regime
with numerical simulation of Eq.~(\ref{radial-GP.equ}).
Polariton fields with $|m| \leq 8$ are included in the simulations,
leading to numerically converged results for the cases considered here.
The desired pumped polariton component
is essentially uniform with a value denoted by $|\bar{\psi}^{+}_{p}|$, except close to R, where it drops to zero.
We design a source that creates such a target function in the absence of the OAM instability.
Once the additional OAM modes are created, the pumped polaritons reach a new steady $\psi^{+}_{m_p} (r)$ that deviates slightly from the
target function. These functions, together with details for the source, are given in the Supplementary Material.
In the simulation,
the stationary states of the nonlinear system are reached after
approximately 0.5ns, and for all results discussed below the state reached after the instability remains unchanged for at least 1ns.

Figure \ref{fig:bargraphintensity} shows the fractions $N_m$ of different OAM modes, defined by
$N_m = \int dr r |\psi^{+}_{m} (r)|^2 / \sum_{m'} \int dr r |\psi^{+}_{m'} (r)|^2$,
in the final steady state reached by the system.
 In Fig.~\ref{fig:bargraphintensity}(a), where $m_p = 0$,
 OAM modes with $m_1=+2$ and $m_2=-2$
are created. These new modes contribute a substantial amount to the total polariton number, here about $10\%$ each.
A different set of parameters yields OAM modes with $m_1=+4$ and $m_2=-4$, Fig.\ \ref{fig:bargraphintensity}(b), which is consistent with the LSA (Fig.~\ref{fig:LSA15}).
Again, the density fractions of  these modes are substantial.
 The creation of polariton OAM modes is not limited to $m_p = 0$.
In Fig.~\ref{fig:bargraphintensity}(c) we demonstrate a case with $m_p = 1$,
%
%
where OAM modes  $m_1=6$ and $m_2=-4$ are created.

We show the corresponding steady state polariton densities, $| \Psi^{+} ( {\rm r})|^2$, in Fig.~\ref{fig:bargraphintensity}  for the three parameter sets discussed above.
For $m_p = 0$ the pump alone would yield a circularly symmetric pattern.
For the first case, where $m = \pm 2$ OAM modes are spontaneously created, the three modes superpose to create a four-fold pattern, and the intensity in Fig.~\ref{fig:bargraphintensity}(d) shows two brighter spots/areas on the x-axis and two dimmer spots/areas on the y-axis. Similarly for the second case, where $m = \pm 4$  OAM modes are created, the three modes ($m = 0 , \pm 4$) generate an eight-fold pattern, with four brighter spots/areas on the x and y axes, and four dimmer (hardly recognizable) spots/areas on the two diagonals, see Fig.~\ref{fig:bargraphintensity}(e). When $m_p = 1$, the spontaneously created modes  are $m_1=6$ and $m_2=-4$, and a simple analytical model for the expected angle dependence of the intensity yields a dependence
 $ \propto \cos ( 5 \phi + \phi_{\rm off} )$, $\phi_{\rm off}$ being an offset angle, in agreement with Fig.~\ref{fig:bargraphintensity}(f).

\begin{figure}
\centerline{\includegraphics[scale=0.85,angle=-00,trim=00 0 20 10,clip=true]{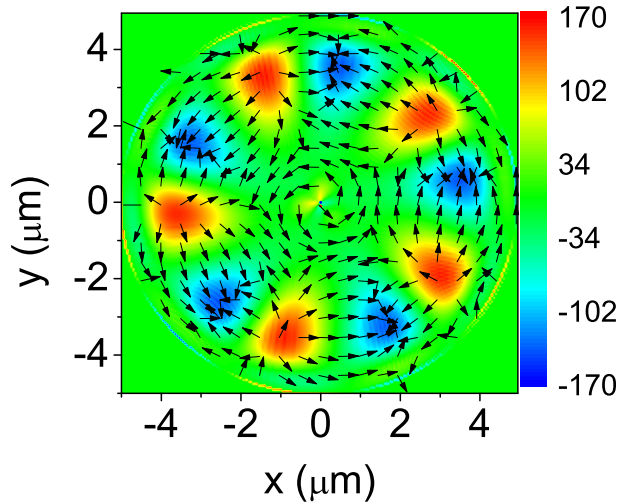}}
\caption{
 The real-space flow pattern given by the normalized current density $\textbf{j}(\textbf{r}) /  | \textbf{j}(\textbf{r}) |$
   (arrows) and the divergence of the current, $ \vec{\nabla} \cdot \textbf{j}(\textbf{r})$,
   shown as color plot, units of $\mu$m$^{-1}$  ps$^{-1}$, for the case of Fig.\ \protect\ref{fig:bargraphintensity} (c),(f). The circulation at the origin, here trivially generated by the $m_p = 1$ pump, is surrounded by five
    generation-annihilation
    pairs
    that emerge as part of the self-sustained optical parametric oscillation.
 For additional flow patterns see Supplement.
  }
\label{fig:current}
\end{figure}

The spontaneous formation of new OAM states through rotational FWM instabilities
also results in the formation of non-trivial or exotic flow patterns, as determined by the current density
$
\textbf{j} = \frac{\hbar}{2 M i}\left(  \Psi^{ \ast}  \vec{\nabla}  \Psi -   \Psi   \vec{\nabla} \Psi^{\ast}  \right) .
$
An example of such a flow pattern is shown in Fig.\ \ref{fig:current} for the dominating component $\Psi = \Psi^+$.
Quite generally, the resulting flow patterns feature spatially distributed generation-annihilation pairs
(i.e.\ sources and sinks of the polariton density corresponding to opposite signs of
 $ \vec{\nabla} \cdot \textbf{j}(\textbf{r})$).
  Figure \ref{fig:current} shows the case of $m_p = 1$, in which the circulation at the origin, which here is trivially generated by the pump, is surrounded by five generation-annihilation pairs.
 One can see that the sources correspond to the intensity peaks in Fig.\ \ref{fig:bargraphintensity}(f)).

In conclusion, the potential for using four-wave mixing instability in semiconductor microcavities as an effective way to spontaneously generate sizable OAM components of light is theoretically demonstrated. Our theory explains the underlying physics and shows how specific individual OAM modes can be created by varying the optical pump parameters. In the future, more general  numerical search algorithms may discover a larger variety of spontaneously formed OAM modes, possibly including non-trivial effects due to polaritonic spin-orbit interaction. Furthermore, other
pump polarizations, taken to be circularly polarized here, and confinement potentials (here step-like) promise to yield yet additional sets of spontaneously formed OAM modes.
We hope that, in the long term, our approach can even yield a device that delivers ``OAM modes on demand.''

We gratefully acknowledge
helpful discussions with Ewan Wright and
financial support from the US NSF under grant ECCS-1406673, TRIF SEOS, and the German DFG
(TRR142, SCHU 1980/5, Heisenberg program).


\begin{thebibliography}{10}

\bibitem{allen-etal.03}
L. Allen, S.~M. Barnett, and M.~J. Padgett, {\em {Optical Angular Momentum}}
  (Institute of Physics Publishing, Bristol, UK, 2003).

\bibitem{torres-torner.11}
J.~P. Torres and L. Torner, {\em {Twisted Photons: Applications of Light with
  Orbital Angular Momentum}} (Wiley-VCH, Singapore, 2011).

\bibitem{mansuripur.17}
M. Mansuripur,  in {\em {Roadmap on Structured Light}} ({J. Opt. {\bf 19}}, pp.
  013001, 2017).

\bibitem{bialynicki-birula-etal.97}
I. Bialynicki-Birula and Z. Bialynicki-Birula, Phys. Rev. Lett. {\bf 78},  2539
   (1997).

\bibitem{courtial-etal.98}
J. Courtial {\it et~al.}, Phys. Rev. Lett. {\bf 81},  4828  (1998).

\bibitem{swartzlander.07}
G. Swartzlander, Phys. Rev. Lett. {\bf 99},  163901   (2007).

\bibitem{courtial-etal.97}
J. Courtial, K. Dholakia, L. Allen, and M.~J. Padgett, Physical Review A {\bf
  56},  4193  (1997).

\bibitem{firth-skryabin.97}
W.~J. Firth and D.~V. Skryabin, Phys. Rev. Lett. {\bf 79},  2450  (1997).

\bibitem{martinelli-etal.04}
M. Martinelli, J.~A.~O. Huguenin, P. Nussenzveig, and A.~Z. Khoury, Phys. Rev.
  A {\bf 70},  013812  (2004).

\bibitem{wang-etal.12}
J. Wang {\it et~al.}, Nature Photonics {\bf 6},  488   (2012).

\bibitem{mair-etal.01}
A. Mair, A. Vaziri, G. Weihs, and A. Zeilinger, Nature {\bf 412},  313  (2001).

\bibitem{loussert-etal.13}
C. Loussert, U. Delabre, and E. Brasselet, Phys. Rev. Lett. {\bf 111},  037802
   (2013).

\bibitem{Savvidis2000}
P.~G. Savvidis {\it et~al.}, Phys. Rev. Lett. {\bf 84},  1547  (2000).

\bibitem{deng-etal.10}
H. Deng, H. Haug, and Y. Yamamoto, Rev. Mod. Phys. {\bf 83},  1489  (2010).

\bibitem{snoke-littlewood.10}
D. Snoke and P. Littlewood, Physics Today {\bf 63},  42  (2010).

\bibitem{turing52}
A. Turing, {Phil. Trans. R. Soc. Lond. B} {\bf 237},  37   (1952).

\bibitem{ball99}
P. Ball, {\em {The self-made tapestry: Pattern formation in nature}} ({Oxford
  University Press}, {New York}, 1999).

\bibitem{staliunas98}
K. Staliunas, Phys. Rev. Lett. {\bf 81},  81  (1998).

\bibitem{Kheradmand2008}
R. Kheradmand {\it et~al.}, Eur. Phys. J. D {\bf 47},  107  (2008).

\bibitem{Dawes2005}
A.~M.~C. Dawes, L. Illing, S.~M. Clark, and D.~J. Gauthier, Science {\bf 308},
  672  (2005).

\bibitem{ardizzone-etal.13}
V. Ardizzone {\it et~al.}, Scientific Reports {\bf 3},  3016  (2013).

\bibitem{weiler-etal.08}
C.~N. Weiler {\it et~al.}, Nature {\bf 455},  07334   (2008).

\bibitem{keeling-berloff.08}
J. Keeling and N.~G. Berloff, Phys. Rev. Lett. {\bf 100},  250401   (2008).

\bibitem{dominici-etal.15}
L. Dominici {\it et~al.}, Science Advances {\bf 1},  1500808   (2015).

\bibitem{newell-etal.93}
A. Newell, T. Passot, and J. Lega, Annu. Rev. Fluid Mech. {\bf 25},  399
  (1993).

\bibitem{krizhanovskii-etal.10}
D.~N. Krizhanovskii {\it et~al.}, Phys. Rev. Lett. {\bf 104},  126402  (2010).

\bibitem{tosi-etal.11}
G. Tosi {\it et~al.}, Phys. Ref. Lett. {\bf 107},  036401  (2011).

\bibitem{leyder-etal.07}
C. Leyder {\it et~al.}, Nature Physics {\bf 3},  628   (2007).

\bibitem{lafont-etal.17}
O. Lafont {\it et~al.}, Appl. Phys. Lett. {\bf 110},  061108   (2017).

\end{thebibliography}
\end{document}